\documentclass[sigconf]{acmart}
\usepackage{makecell}
\usepackage{makecell}
\usepackage{balance}
\AtBeginDocument{%
  }

\copyrightyear{2026}
\acmYear{2026}
\setcopyright{cc}
\setcctype{by}
\acmConference[CIKM '26]{Proceedings of the 35th ACM International Conference on Information and Knowledge Management}{November 07--11, 2026}{Rome, Italy}
\acmBooktitle{Proceedings of the 35th ACM International Conference on Information and Knowledge Management (CIKM '26), November 07--11, 2026, Rome, Italy}
\acmDOI{10.1145/3799682.3840102}
\acmISBN{979-8-4007-2539-5/2026/11}

\newcommand{\systemname}{\textit{Feed SR}}

\begin{document}


\title{An Industrial-Scale Sequential Recommender for LinkedIn Feed Ranking}

\author{Lars Hertel}\authornote{Work done while at LinkedIn.}\authornote{Both authors contributed equally to this research.}
\author{Gaurav Srivastava}\authornotemark[2]
\email{gsrivastava@linkedin.com}

\author{Syed Ali Naqvi}
\author{Satyam Kumar}
\author{Yue Zhang}\authornotemark[1]
\author{Borja Ocejo}\authornotemark[1]

\affiliation{%
  \institution{LinkedIn}
  \country{Mountain View, CA, USA}
}

\author{Benjamin Zelditch}
\email{bzelditch@linkedin.com}
\author{Adrian Englhardt}
\author{Hailing Cheng}\authornotemark[1]
\author{Andy Hu}
\author{Antonio Alonso}
\author{Daming Li}

\affiliation{%
  \institution{LinkedIn}
  \country{Mountain View, CA, USA}
}


\author{Siddharth Dangi}
\email{sdangi@linkedin.com}
\author{Chen Zhu}
\author{Mingzhou Zhou}\authornotemark[1]
\author{Wanning Li}
\author{Tao Huang}
\author{Fedor Borisyuk}
\affiliation{%
  \institution{LinkedIn}
  \country{Mountain View, CA, USA}
}

\author{Ganesh Parameswan}
\email{gaparameswaran@linkedin.com}
\author{Birjodh Tiwana}
\author{Sriram Sankar}\authornotemark[1]
\author{Qing Lan}
\author{Julie Choi}
\author{Souvik Ghosh}\authornotemark[1]

\affiliation{%
  \institution{LinkedIn}
  \country{Mountain View, CA, USA}
}

\renewcommand{\shortauthors}{Lars Hertel et al.}



\renewcommand{\shortauthors}{Lars Hertel et al.}


\begin{abstract}
LinkedIn Feed enables professionals worldwide to discover relevant content, build connections, and share knowledge at scale. We present \emph{Feed Sequential Recommender} ({\systemname}), a transformer-based sequential ranking model for LinkedIn Feed that replaces a DCNv2-based ranker and meets strict production constraints. We detail the modeling choices, training techniques, and serving optimizations that enable deployment at a scale of 1.2 billion members. {\systemname} has been serving the majority of LinkedIn’s Feed traffic for over three months and shows significant improvements in member engagement (+2.10\% time spent, +3.52\% like, comments, or reshares) in online A/B tests compared to the existing production model. We also describe our deployment experience with alternative sequential and LLM-based ranking architectures and why {\systemname} provided the best combination of online metrics and production efficiency.

\end{abstract}

\ccsdesc[500]{Information systems~Recommender systems}
\ccsdesc[500]{Information systems~Learning to rank}
\ccsdesc[300]{Information systems~Personalization}

\keywords{Sequential Recommendation, Generative Recommendation, Recommender systems, Ranking}

\maketitle

\begingroup
\renewcommand{\thefootnote}{\fnsymbol{footnote}}
\footnotetext[1]{Both authors contributed equally to this research.}
\footnotetext[2]{Work done while at LinkedIn.}
\endgroup

\section{Introduction}
LinkedIn Feed enables professionals to share knowledge and ideas through text, images, and video. Members viewing the Feed are presented with posts from their network, as well as relevant posts from members they are not connected to. In addition, the Feed includes other content such as job changes, job opportunities, and suggested connections. Posts in the Feed are ranked by predicted relevance — the likelihood that a member will take actions such as clicking, liking, commenting, or sharing — and improving prediction accuracy drives growth for both the LinkedIn Feed and the broader LinkedIn ecosystem. Sequential recommendation methods have shown promise in improving prediction accuracy of member actions.

However, deploying these methods in LinkedIn Feed presents intertwined challenges: 1.2B+ members with a highly long-tailed activity distribution, a dynamic corpus where newly created posts can accumulate engagement rapidly, and strict real-time constraints (a few hundred milliseconds latency and tens of thousands of QPS throughput). Motivated by these challenges, we present {\systemname}, a system–model co-design for deploying sequential recommendation in LinkedIn Feed ranking. Our contributions are as follows:
\begin{itemize}
    \item \textbf{Production deployment at LinkedIn scale.}
    {\systemname} replaces the production DCNv2 ranker for LinkedIn Feed's 1.2B+ members, yielding +2.10\% time spent and +3.52\% likes, comments, and reshares in online A/B tests (Section~\ref{sec:online-results}).

    \item \textbf{System design under real-time ranking constraints.}
    We deploy sequential ranking with a disaggregated CPU/GPU serving stack, shared-context batching for hundreds of candidates per request, zero-copy Arrow-to-tensor transfer, and \emph{SRMIS}, a custom Flash-Attention kernel with $\sim 2\times$ speedup over PyTorch SDPA (Section~\ref{sec:online-serving}).

    \item \textbf{Production-driven modeling techniques.}
    We introduce late fusion for candidate/context features, LLM-derived member profile embeddings, and within-session randomization to address feature expressivity, sparse-history members, and train--serve skew from intra-session label correlations (Sections~\ref{sec:late-fusion},~\ref{sec:member-profile-embed},~\ref{sec:in-session-leakage}).

    \item \textbf{Deployment lessons from alternative architectures.}
    We report lessons from production-scale experiments with fine-tuned LLM-Rankers and TransAct-style~\cite{xia2023transact} history encoders that motivated {\systemname}'s final quality--latency--efficiency trade-off (Section~\ref{sec:alternative-approaches}).
\end{itemize}

\section{Related Work}
\label{sec:related_work}

Our work intersects three active research directions: industrial recommendation models with rich feature sets, sequential recommendation with long user histories, and generative recommendation models that amortize computation across multiple candidate scores.

\paragraph{Industrial Recommendation Models.}
Deep Learning Recommendation Models (DLRMs) combine dense numerical features, sparse categorical IDs, and cross features to deliver high-quality ranking in production systems \cite{naumov2019dlrm,cheng2016wide}. Enhancements such as explicit cross layers (e.g., DCNv2) and multi-task sharing (e.g., MMoE) provide practical performance gains in large-scale ranking \cite{wang2021dcnv2,ma2018mmoe}. These models remain dominant in online settings due to their ability to integrate rich feature interactions.

\paragraph{Sequential Recommendation.}
Modeling user behavior sequences is essential for capturing temporal patterns and long-range interests. Transformer–based sequence encoders like SASRec and bidirectional masked pretraining such as BERT4Rec represent the state-of-the-art for next-item prediction \cite{kang2018sasrec,sun2019bert4rec}. However, these do not directly apply to action prediction as used in ranking settings. Industrial variants extend these ideas with target-aware attention and behavior-sequence modeling, including TransAct~\cite{xia2023transact}, Behavioral Sequence Transformer~\cite{chen2019behavior}, and Pinnerformer~\cite{pancha2022pinnerformer}. We compare against TransAct-style encoders in Section~\ref{sec:alternative-approaches}.

\paragraph{Generative Recommendation Models.}
Generative recommendation models (GRMs) treat user histories as token sequences and use transformer transduction for next-item prediction, enabling amortized scoring across many targets \cite{zhai2024actions}. HSTU-style architectures have demonstrated competitive quality with improved efficiency for long histories, but their integration with production-level feature stacks is non-trivial. Recent industrial GRM frameworks such as MTGR emphasize retaining engineered features and optimizing system deployment for ranking at scale \cite{han2025mtgr}. OneRec similarly explores iterative preference alignment in a unified generative retrieve-and-rank framework \cite{deng2025onerec}.

Our work extends these lines of work by integrating a sequential model with rich production feature sets under strict latency and throughput constraints in LinkedIn Feed ranking, while retaining high offline and online quality.

\section{Background}
\subsection{LinkedIn Feed}
LinkedIn's mission is to connect the world’s professionals to make them more productive and successful. As part of this mission, LinkedIn Feed serves to provide access to ideas, learning, and inspiration from members' networks and beyond. To achieve this, the Feed system optimizes for a variety of actions, among them:
\begin{enumerate}
    \item Long Dwell: dwelling on a post longer than a specified threshold of time, depending on the post-type
    \item Contribution: a like, comment, or share on a post.
\end{enumerate}
To generate a list of posts for a member, candidate posts are first retrieved from multiple sources \cite{ramanujam2025large,Ghike_Gupta_2016} and combined. Then a ranking model predicts the likelihood of each action. The resulting scores are combined via weights in an objective function that is used to rank the posts. Finally, additional business logic is applied on the ranked post list.

\subsection{Feed ranking model}\label{sec:existing}
The existing Feed ranking model uses post impressions as training samples. The labels correspond to the possible actions (e.g. long-dwell, like, comment, share, etc.). Impressions that do not have a click are retained randomly with 0.1 probability. The model is initially trained on three weeks of Feed data, and then incrementally trained on a daily basis. The ranking model uses a large number of features that can be divided into numeric features, content embeddings, ID embeddings, and categorical features. The model is implemented in Tensorflow and uses DCNv2 with multi-task output layers (see \cite{borisyuk2024lirank} for full details).

Key online metrics are \emph{time spent} and the volume of \emph{likes, comments, and reshares}.

\section{{\systemname} -- Experiments and offline results}

\begin{figure}
  \centering
  \includegraphics[width=\linewidth]{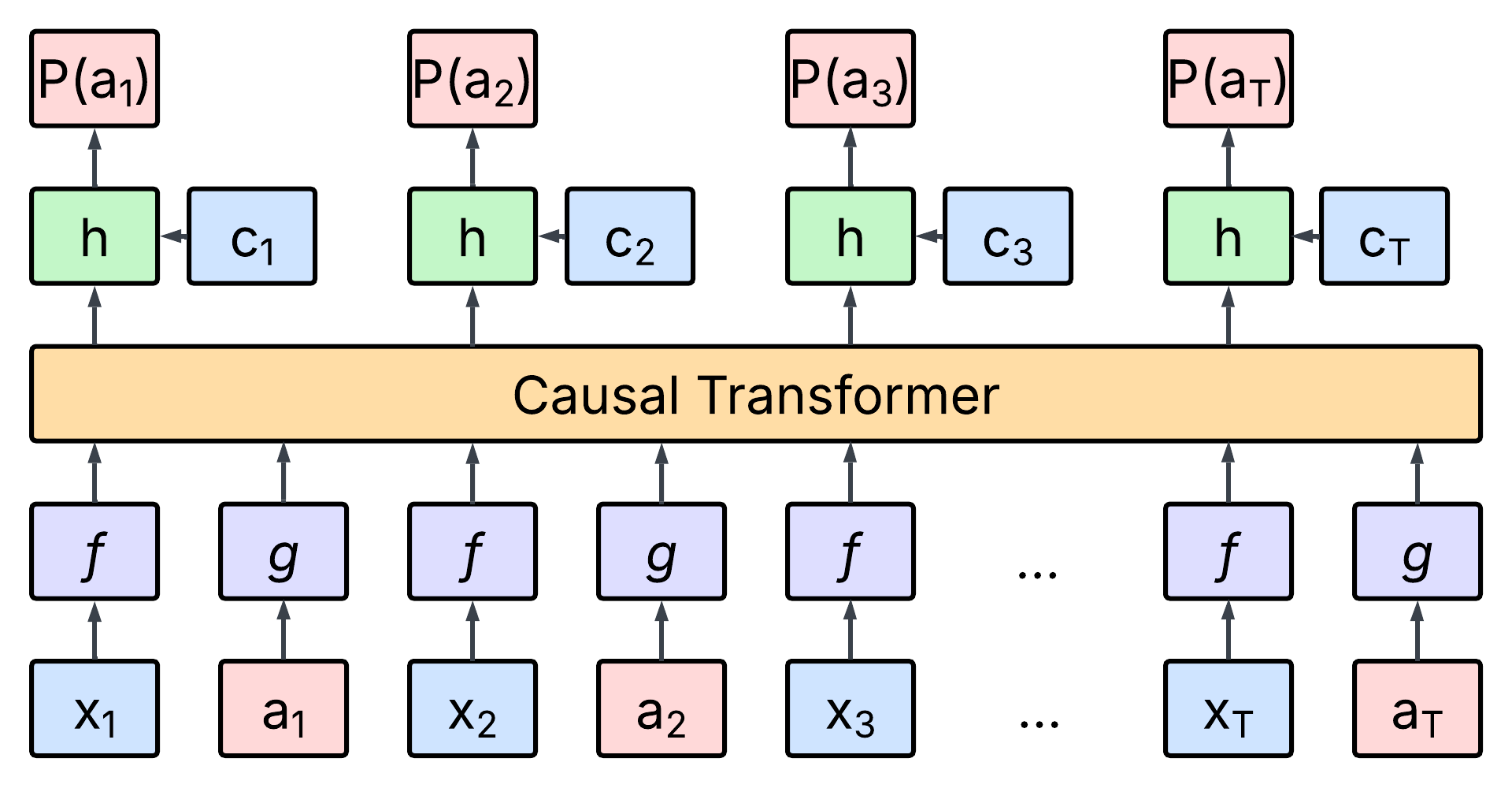}
  \caption{The {\systemname} model architecture.}
  \Description{Input is a sequence of Feed post and action embedding, interleaved.}
  \label{fig:architecture}
\end{figure}

\subsection{Architecture overview}\label{sec:architecture}
{\systemname} is a new ranking system for the LinkedIn Feed based on sequential recommendation. The model architecture is based on the ranking model in \cite{zhai2024actions} and enhanced to fit LinkedIn Feed's specific use case, product requirements, and infrastructure. Figure~\ref{fig:architecture} shows a diagram of the model in which we interleave posts with actions similar to \cite{zhai2024actions}. The sequence of interleaved posts and actions are processed by a number of causal transformer blocks \cite{vaswani2017attention} (details in Section~\ref{sec:transformer-arch}). After the transformer blocks, outputs corresponding to interleaved action inputs are discarded as shown in Figure~\ref{fig:architecture}~\cite{zhai2024actions}. The result is concatenated with additional context features (Section~\ref{sec:late-fusion}) and further processed by a "head-architecture" deep neural network with a multi-task output (Section~\ref{sec:head-arch}). The multi-task output predicts the actions that the member took on the post. The model is trained on member interaction histories to predict actions at all points in the interaction sequence. The loss function is a binary cross-entropy loss and is applied to all sequence positions. During inference, the candidates to be ranked are appended to the end of the sequence and scored at once. More details on inference can be found in Section~\ref{sec:online-gpu-optimizations}.

\subsection{Architecture Details}

\subsubsection{Model Input}\label{sec:model-input}

The model input consists of a history of impressed Feed posts interleaved with the actions taken by the member on each post. Histories are formed from one year of training data (defined in Section~\ref{sec:existing}). For each member, we keep the most recent $T{=}1000$ impressions, ordered chronologically from oldest to newest. Due to negative sampling, $T{=}1000$ impressions cover a sufficiently long time horizon, even for very active members. Furthermore, recency tends to be an important factor for selecting data. A post at sequence position $t$ is represented by $J$ features $\{x_{t,1}, \dots, x_{t,J}\}$. Each feature gets encoded by a feature-specific transform $f_j$ such as embedding lookup, log transform, or identity. The resulting post representation is
\begin{equation}
X_t = \operatorname{concat}\bigl(f_1(x_{t,1}), \dots, f_J(x_{t,J})\bigr)
\in R^{d_{\text{seq}}}.
\label{eq:xt-concat}
\end{equation}
We denote the stacked sequence of post representations by $X_{\text{seq}} = [X_1, X_2, \ldots, X_T]$. Similarly, given a multi-hot action vector for $M$ actions $a_t \in \{0, 1\}^{M}$, we form an action representation $A_t = a_t W_a + b_a\in R^{d_{\text{seq}}}$ using a learnable projection. The stacked action sequence is $A_{\text{seq}} = [A_1, \ldots, A_T]$. Finally, the $2T$-length transformer input is constructed by interleaving post representations and action representations:
\begin{equation}
X_{\text{in}} = [ X_1, A_1, X_2, A_2, \ldots, X_T, A_T ].
\label{eq:xin}
\end{equation}
In our experiments, we set the transformer embedding dimension to be $d_{\text{model}} = d_{\text{seq}}$.

\subsubsection{Transformer architecture and variations}\label{sec:transformer-arch}
We use a decoder-only transformer with a pre-LayerNorm formulation \cite{xiong2020layer}, rotary positional embeddings (RoPE) \cite{su2024roformer}, and scaled residual connections. The forward pass of the block is defined in Eqs.~\eqref{eq:qkv}--\eqref{eq:resid-ffn}.

\begin{align}
Q,K,V &= W_q\operatorname{LN}(X_{\text{in}}),\; W_k\operatorname{LN}(X_{\text{in}}),\; W_v\operatorname{LN}(X_{\text{in}}) \label{eq:qkv}\\
Q_r,K_r &= \operatorname{RoPE}(Q,K) \label{eq:rope}\\
\mathrm{Attn} &= W_o\,\operatorname{Concat}\!\left(\operatorname{SDPA}(Q_r,K_r,V;\text{causal})\right) \label{eq:attn}\\
Y &= \operatorname{RescaleAndAdd}(X_{\text{in}},\mathrm{Attn}) \label{eq:resid-attn}\\
Z &= \operatorname{RescaleAndAdd}\!\left(Y,\operatorname{FFN}(\operatorname{LN}(Y))\right) \label{eq:resid-ffn}
\end{align}

Equations~\eqref{eq:qkv}--\eqref{eq:attn} compute query, key, value projections, apply RoPE, and perform causal SDPA followed by output projection $W_o$. Equations~\eqref{eq:resid-attn}--\eqref{eq:resid-ffn} apply scaled residual updates similar to \cite{bachlechner2021rezero}. We define
\begin{equation}
\operatorname{RescaleAndAdd}(u,v) \triangleq u + \alpha v,
\end{equation}
where $\alpha$ is a learnable scalar initialized to one; initializing $\alpha$ to zero recovers ReZero~\cite{bachlechner2021rezero}. Finally, the transformer output $Z \in R^{B \times 2T \times d_{\text{model}}}$ is reduced to $R^{B \times T \times d_{\text{model}}}$ by discarding odd output positions which correspond to action inputs similar to \cite{zhai2024actions}. The result is passed to the head architecture together with the context features $X_{\text{context}}$ for late-fusion prediction; see Sections~\ref{sec:late-fusion} and~\ref{sec:head-arch} for details.

We use pre-LayerNorm and discuss its criticality for model stability in Section~\ref{sec:training-stability}. We evaluated different attention activations in Eq.~\eqref{eq:attn}, including Softmax, Sigmoid, SiLU, and ReLU. Contrary to observations in \cite{zhai2024actions}, Softmax activation exceeded the performance of Sigmoid, SiLU, and ReLU in our setting. Performance drop in AUC when using alternative attention activations is presented in Table~\ref{tab:app-arch-ablations}. We use scalar \texttt{RescaleAndAdd} skip connection instead of vanilla residual addition and discuss the importance of different skip connections in Section~\ref{sec:training-stability}

We also evaluated replacing the {\systemname} transformer blocks with HSTU layers~\cite{zhai2024actions}, but observed a consistent performance degradation as evident in Table~\ref{tab:app-arch-ablations}. For instance, at matched compute ($10^{17}$ FLOPs), the Long Dwell AUC decreases by $0.21\%$ with HSTU.

\subsubsection{Late Fusion of Features}\label{sec:late-fusion}
We propose to use \emph{late fusion} for select context features in which we concatenate these features to the transformer output as shown in Figure~\ref{fig:architecture}. Firstly, injecting features after the transformer reduces compute and the need to store these features online in the history. Secondly, historical patterns in numeric features may only be weakly useful, justifying this architecture from a modeling point of view. In offline experiments, moving one-third of late-fused features to early fusion yields only a minor lift ($0.04\%$) in Long Dwell AUC, while late fusion reduces training step time by $12\%$. These results were further confirmed through an online A/B test, which showed parity between late-fusion and early-fusion models on top-line metrics. The features moved to late-fusion are primarily numeric signals that capture item popularity and viewer-author affinity (between the member viewing the Feed and the post's author). Note that we were not able to move all numeric features to late-fusion without significant AUC drops. This may be due to the fact that, while historical patterns of numeric features may themselves not be informative, the features can still be useful in modulating sequence attention.

\subsubsection{Head Architecture}\label{sec:head-arch}
In Section~\ref{sec:transformer-arch} we introduced the {\systemname} head architecture, which generates prediction scores. We experiment with several head architectures: (i) \textbf{Linear} (a single affine layer), (ii) \textbf{MLP} (a three-layer MLP with nonlinearity), (iii) stacked and parallel \textbf{DCNv2}~\cite{wang2021dcn}, and (iv) \textbf{MMoE}~\cite{ma2018modeling}. Table~\ref{tab:app-arch-ablations} summarizes their relative offline AUCs. We observe that the parallel DCNv2 head achieves the best performance among the evaluated heads on both Long Dwell and Contributions responses. The linear head shows the largest degradation, while MLP, MMoE, and stacked DCNv2 all fall within $\sim0.3\%$ of the baseline. These results suggest that some form of nonlinear feature combination at the head is important, while differences among nonlinear architectures are more modest.

\subsection{Features}\label{sec:features}
{\systemname} reduces the number of features compared to the existing production model by about 80\%. Thorough empirical evaluation showed that the removed features did not yield further AUC improvements in {\systemname}. This suggests that the transformer supplied with a raw interaction history can replace many hand-engineered numeric features. However, we also found that some key features could not be replaced. Firstly, item popularity signals remain crucial as {\systemname} cannot capture these signals from the member history (e.g., candidate popularity features yield $+2.5\%$ Long Dwell AUC). Secondly, while viewer-to-author affinity can be captured from the sequence, existing counts of these over longer time windows still provide AUC improvements when included in the model (e.g., removing member-to-member affinity features yields a $0.3\%$ Long Dwell AUC drop). This may be due to the important role that viewer-to-author relations play on the LinkedIn Feed.

  \begin{table}[htbp]
  \centering
  \small
  \begin{tabular}{@{}lcc@{}}
  \toprule
  Experiment & \makecell{Long Dwell\\AUC $\Delta$\%} & \makecell{Contributions\\AUC $\Delta$\%} \\ \midrule
  \makecell[l]{Baseline (RoPE PE, DCNv2-parallel head,\\Softmax attention, rescaling residual, late fusion)} & -- & -- \\ \midrule
  RoPE $\rightarrow$ learned position embedding                  & $-0.19\%$ & $-0.16\%$ \\
  RoPE $\rightarrow$ no position embedding                       & $-0.91\%$ & $-0.48\%$ \\ \midrule
  DCNv2 parallel $\rightarrow$ Linear                            & $-1.20\%$ & $-0.48\%$ \\
  DCNv2 parallel $\rightarrow$ MLP w/ ReLU                       & $-0.13\%$ & $-0.08\%$ \\
  DCNv2 parallel $\rightarrow$ DCNv2 stacked                     & $-0.27\%$ & $-0.19\%$ \\
  DCNv2 parallel $\rightarrow$ MMoE                              & $-0.16\%$ & $-0.12\%$ \\ \midrule
  Softmax attn. $\rightarrow$ SiLU                               & $-0.09\%$ & $-0.05\%$ \\
  Softmax attn. $\rightarrow$ ReLU                               & $-0.25\%$ & $-0.12\%$ \\
  Softmax attn. $\rightarrow$ Sigmoid                            & $-0.15\%$ & $-0.12\%$ \\ \midrule
  Feed SR $\rightarrow$ HSTU                                    & $-0.23\%$ & $-0.28\%$ \\ \midrule
  Scalar rescaling $\rightarrow$ LayerScale                      & $+0.01\%$ & $+0.07\%$ \\
  Scalar rescaling $\rightarrow$ Dense gating$^\dagger$          & $+0.03\%$ & $+0.02\%$ \\
  Scalar rescaling $\rightarrow$ ReZero$^\dagger$                & $-0.09\%$ & $-0.08\%$ \\ \midrule
  Late fusion $\rightarrow$ Early fusion                         & $+0.04\%$ & $+0.01\%$ \\ \bottomrule
  \end{tabular}
  \caption{Architectural ablations for {\systemname}. Relative AUC change against the baseline. $^\dagger$ Reported with LR reduced by half; both diverged at the default learning rate.}
  \label{tab:app-arch-ablations}
  \vspace{-2.5em}
  \end{table}

\subsection{RoPE vs Learned absolute embeddings.}\label{sec:rope}
As mentioned in Section~\ref{sec:transformer-arch}, we use rotary positional embeddings (RoPE)~\cite{su2024roformer} to encode token positions. Prior to RoPE, we experimented with additive learned absolute position embeddings, where each token at absolute sequence index $t$ is assigned a learned vector $p_t$ and the transformer input is formed as $x_t + p_t$. In our setting, learned absolute embeddings resulted in unstable average prediction scores during training which is undesirable. To mitigate this, we replace learned absolute embeddings with RoPE, which encodes position through a shared, deterministic rotation applied to $(Q,K)$, generating $(Q_{r}, K_{r})$, rather than a learned lookup table over absolute indices. We find that RoPE improves score stability, keeping a coefficient of variation for average predicted scores around 1\%, and also yields metric gains over learned absolute position embeddings, as shown in Table~\ref{tab:app-arch-ablations}.

\subsection{Member Profile Embeddings}\label{sec:member-profile-embed}
Member profile embeddings are an LLM-based dense representation that captures comprehensive information from LinkedIn member profiles. These embeddings are generated by aggregating member profile information with a Qwen3 0.6 billion parameter \cite{yang2025qwen3} fine-tuned model. To efficiently incorporate this dense embedding into the model, we integrate it as a late‑fused context feature, avoiding any increase in the transformer’s dimensionality.
Together with item popularity features, member profile embeddings are key signals for members with sparse interaction histories. Embeddings are refreshed daily. This means that as members evolve their profiles, the embeddings stay aligned with each member's current professional interests. Empirically, adding profile embeddings as a late‑fused dense feature improves Long‑Dwell AUC, with more than $+2\%$ AUC gains for members with fewer than 10 historical actions.

\subsection{Training Techniques}

 \subsubsection{Training stability}\label{sec:training-stability}
  Some architectural changes destabilize training: AUC may collapse to 0.5 or drift mid-run. We suspect this stems from statistical heterogeneity across tokens --- tokens at different positions correspond to different items, actions, and time gaps --- which may stress gradient flow through the residual stream. The following techniques helped stabilize training:
  \begin{itemize}
      \item \textbf{Pre-LN is essential}; without it, AUC collapses to 0.5.
       \item \textbf{A stable residual / skip-connection}: vanilla residual addition was intermittently unstable, while dense gating~\cite{chai2020highway}, LayerScale~\cite{touvron2021going}, ReZero~\cite{bachlechner2021rezero}, and scalar rescaling (Section~\ref{sec:transformer-arch}) all stabilized training with marginal AUC differences (Table~\ref{tab:app-arch-ablations}); dense gating and ReZero additionally required reducing the learning rate.
      \item \textbf{Learning-rate reduction} stabilized training in most cases; when insufficient, combining it with \textbf{dense gating} was the broadest-applicable recipe, at the cost of additional parameters and memory.
  \end{itemize}

\subsubsection{Incremental Training}\label{sec:incremental-training}
Incremental training is a core component of LinkedIn’s online recommendation system and has proven to be both effective and essential for large-scale social media platforms. LinkedIn Feed’s ranking models are updated daily using newly arrived interaction data, and {\systemname} follows the same paradigm. During cold-start training, the loss is computed over the full interaction history, as described in Section~\ref{sec:architecture}. During incremental updates, we compute the loss only on newly observed interactions, while still providing the full historical sequence as input to the model. Although daily data is dominated by highly active members, we have not observed evidence so far that this incremental training strategy degrades model quality for less frequent members. Early online experiments at low traffic percentages show additional metric gains beyond those reported in Section~\ref{sec:online-results}, which excludes incremental training.

\subsubsection{In-Session Leakage}\label{sec:in-session-leakage}
The {\systemname} input sequences often contain multiple items from the same session. This poses a problem, as labels such as long-dwell are correlated within a session. For example, conditioning on the first item in a session having dwell time $>15$s increases the likelihood that subsequent items in the same session also exceed 15s, relative to items outside the session. The result is a train--serve discrepancy: during training, the model can use previous in-session labels; at serving time, those labels are not yet available. To mitigate in-session leakage when constructing training sequences, we experimented with (1) randomizing the order of items within each session and (2) masking items from the same session during attention. While the second approach is more principled, we found that both methods resolved the observed overfitting issues. Furthermore, (2) resulted in slower training due to the data-dependent mask creation. We therefore use simple randomization within the session in the final model.

\subsubsection{Training details}
The {\systemname} incremental training infrastructure is orchestrated through Apache Airflow and Flyte. Member histories are generated in daily increments and used for training and serving. Both cold-start (CS) and warm-start (WS) training use a global batch size of 1024 and the AdamW optimizer \cite{loshchilov2017decoupled}. CS training is trained using 16 H200’s and WS uses 8 H200’s. CS trains using OneCycleLR while WS training uses the final CS learning rate.

\subsection{Scaling Laws}\label{sec:scaling-laws}

\begin{figure}
    \centering
    \includegraphics[width=0.9\linewidth]{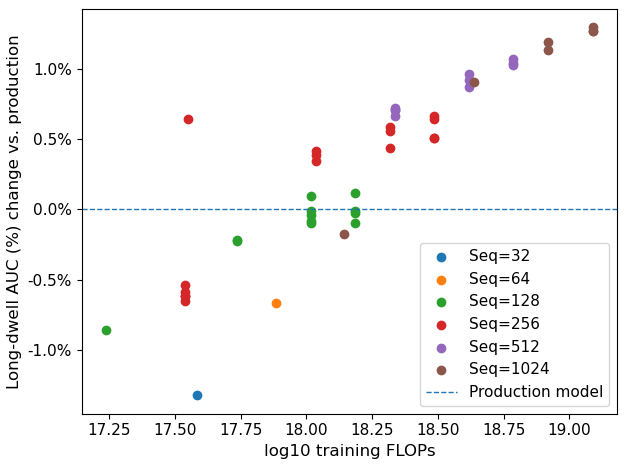}
    \caption{Scaling of Long Dwell AUC as a function of training FLOPs for {\systemname}. Baseline is the current Feed production model.}
    \label{fig:scaling-long-dwell}
    \centering
\end{figure}

 We perform experiments to observe scaling behavior in {\systemname}, sweeping sequence lengths in $\{32, 64, 128, 256, 512, 1024\}$, transformer depths in $\{1, 4, 8, 12\}$, and ID-embedding dimensions in $\{16, 32, 64\}$, spanning approximately $10^{17}$ to $10^{19}$ training FLOPs. Figure~\ref{fig:scaling-long-dwell} shows how Long Dwell AUC scales with log of the training FLOPs. For every order of magnitude increase in training FLOPs, the evaluation Long Dwell AUC improves by approximately 0.0093, demonstrating a consistent positive scaling effect. Training FLOPs are an approximate function of the number of dense parameters times the sequence length~\cite{casson2023transformerflops}. We also performed ablations that scale model capacity along a single axis at a time: sequence length, number of transformer layers, and ID embedding dimension. Among these, increasing sequence length exhibits the most consistent scaling behavior across metrics (including Long Dwell AUC and Contributions AUC). This is intuitive, as increasing sequence length simultaneously increases the number of training examples and the signal per example. In contrast, increasing transformer depth or embedding dimension adds representational capacity without introducing new information. Overall, these results suggest that to obtain reliable scaling laws across objectives, capacity should be scaled jointly across multiple dimensions rather than by increasing a single dimension in isolation.

\section{Alternative Approaches Considered}\label{sec:alternative-approaches}
\subsection{LLM-Ranker}
Before {\systemname}, we explored an LLM-Ranker system in which all the features of a candidate post were represented as text and passed into an LLM as part of a prompt. Given an input prompt with a particular structure and ending with a question such as "Will the member click on this post?", the LLM was fine-tuned to output "Yes" or "No" as the next token. The LLM's output logit for "Yes" was then extracted and interpreted to be the model's prediction for P(click). The same process was used to make predictions for other probabilities of interest (P(like), P(comment), etc.).

This LLM-Ranker showed promising offline results in early experiments and presented various appealing properties. For instance, because the LLM-Ranker was initialized from an already pre-trained LLM (e.g., LLama 3), it already contained a lot of "world knowledge". Furthermore, as the LLM internally also had a transformer-backbone, it also benefited from sequential training and multi-item scoring like {\systemname}.

However, the LLM-Ranker also had several key disadvantages:
\begin{itemize}
    \item Although certain features were natural to represent as text in the input prompt, it was difficult to encode numeric features as text (e.g., "\# of Likes: 382"), and the LLM was not always able to leverage them well to improve the model's prediction performance.
    \item Because it took hundreds of tokens to represent each post, encoding a member's post interaction history took tens of thousands of tokens, making the model expensive to train and serve due to the large sequence input being passed into the transformer. In contrast, although {\systemname} also processes a member's interaction history with a transformer, each post in the history is represented only by 2 tokens (item and action embeddings), which is much more efficient.
    \item The LLM-Ranker never achieved superior online performance over the existing production model. Although it performed well on posts from out-of-network recommendations, the model struggled with network-based recommendations, because it was difficult to encode the strength of network relationships in a text prompt.
\end{itemize}
{\systemname} addresses these limitations through explicit handling of popularity and affinity signals via late fusion, compact post representations that enable efficient serving over long interaction histories, and ID-based embeddings that capture strong network relationships.

\subsection{TransAct}
We also experimented with augmenting the existing production ranker with history encoding methods such as TransAct \cite{xia2023transact}, Behavioral Sequence Transformer (BST) \cite{chen2019behavior}, and Deep Interest Network (DIN) \cite{zhou2018deep}. TransAct improved offline and online metrics. However, it also resulted in significant increases in training time and inference latency, especially for longer sequences. While we prototyped the usage of multi-item scoring in TransAct \cite{hertel2024efficient}, the method was difficult to deploy in a stack implemented for point-wise scoring. Making the sequence the central part of {\systemname} naturally resolved these challenges by amortizing sequence computation during training and inference.

\section{System architecture}
\subsection{Overview}
\begin{figure}
    \centering
    \includegraphics[width=0.8\linewidth]{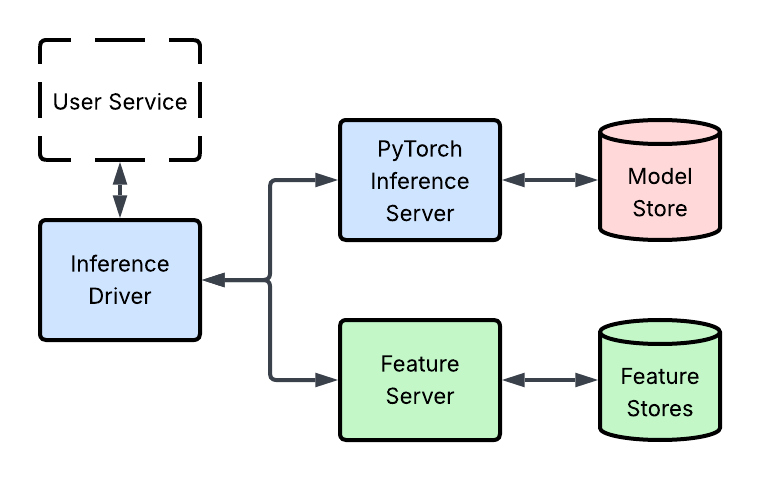}
\caption{System Architecture of {\systemname}}
    \Description{A diagram showing member history venice store, data pipelines, lix setup, feature transformation, {\systemname} service, candidate feature extraction and transformation, score generation.}
    \label{fig:system-architecture}
\end{figure}

Figure~\ref{fig:system-architecture} shows a high level architecture diagram of the inference infrastructure that powers our {\systemname} system.
Our inference system employs a disaggregated architecture that separates CPU and GPU workloads to enable independent scaling and optimal resource utilization.

\subsubsection{Disaggregated Inference Components}
The inference driver is a CPU-based service that handles feature fetching, feature tracking, and request-context-specific CPU-bound transformations. These operations are not GPU-friendly and are therefore isolated from GPU resources. The PyTorch inference server is a Python-based service optimized for GPU execution. It exposes a high-performance gRPC interface that wraps Apache Arrow buffers inside protobuf messages, enabling zero-copy conversion of large data payloads directly to PyTorch tensors. This design eliminates expensive serialization and memory copy operations, allowing the transformer model to process features efficiently on the GPU.

\subsubsection{Efficient Feature Serving}
Our feature serving infrastructure minimizes inference latency through precomputation and efficient data layouts. Member history features are generated offline and stored as compact Arrow columnar buffers in key-value stores, enabling efficient access with low memory overhead. Document features are handled similarly and fetched at serving time.

During inference, the driver retrieves and joins member and document features, then transmits the combined feature set as Arrow bytes to the PyTorch inference server. The system relies on zero-copy conversion between Arrow buffers and PyTorch tensors, which is especially important for large history features. Both services are implemented in Python but restricted to high-performance libraries (NumPy, PyTorch, Arrow) to minimize overhead and maximize throughput.

\subsection{Improving latency and QPS}\label{sec:online-serving}

\subsubsection{CPU-Side Optimization}
We optimized the inference pipeline to eliminate CPU-bound data processing bottlenecks. Member history parsing improved from 450 ms to 2 ms (225$\times$), while sparse-to-dense tensor conversion dropped from 254 ms to 5 ms per feature (50$\times$), directly increasing production throughput.

Member history parsing was vectorized using NumPy strided arrays for zero-copy, bulk processing, replacing item-centric loops with feature-centric access and reducing complexity from $O(N\times F)$ to $O(F)$. Sparse-to-dense conversion similarly replaced nested Python loops with PyTorch tensor indexing and batched assignments, reducing complexity from $O(N\times M\times D)$ to $O(N+M\times D)$.

Together, these changes reduced CPU cycles by 66\%, instructions by 71\%, cache misses by 90\%, and branch mispredictions by 72\% relative to the original implementation, enabling efficient scaling under high-throughput, real-time inference workloads.

\subsubsection{GPU-Side Optimization}\label{sec:online-gpu-optimizations}
\paragraph{Shared Context Batching}
{\systemname} scores $N$ candidates (typically 512) per request, all sharing the same interaction history. We therefore batch candidates into a single forward pass using a custom attention mask, reducing redundant computation and achieving an 80$\times$ transformer speedup for typical workloads (history length 1000, $\sim$500 candidates).

\paragraph{Custom Flash Attention Kernel}
PyTorch's SDPA falls back to a non-Flash Attention implementation when custom masks are provided. We developed \textit{SRMIS}, a specialized CUDA kernel that extends Flash Attention to support the multi-item scoring pattern. It encodes the attention pattern using only two scalar parameters, \texttt{context\_length} and \texttt{candidate\_length}: context tokens attend causally to preceding positions, while candidates attend only to the full context and themselves. This avoids materializing $O((L+N)^2)$ attention masks, skips tiles outside the valid attention region, and retains Flash Attention's tiled computation and online softmax, yielding an average 2$\times$ speedup over masked SDPA.

\subsection{Model Training Optimizations}\label{sec:training-optimizations}

To improve the efficiency of training for {\systemname}, the following key optimizations were implemented. These strategies significantly reduced training time, computational overhead, and resource usage.

\begin{table}[ht]
\centering
\small 
\begin{tabular}{lccc}
\hline
\textbf{Optimization Applied} & \textbf{e2e GPU Hours Reduction}\\ \hline
Efficient Metrics Computation Kernel & 22\% \\
Optimizer Fusing and Gradient Scaling & 15\% \\
Fused Data Loading and Processing & 50\% \\
Parallelized Evaluation & 16\% \\
\hline
\end{tabular}
\caption{Training performance relative improvements}
\label{table:Training_optimizations_table}
\vspace{-2.5em}
\end{table}

\subsubsection{Efficient Metrics Computation Kernel}
The original multilabel AUC computation was performed sequentially, incurring substantial overhead ($\sim66$ ms per step) due to masked confusion-matrix operations and frequent \texttt{cudaStreamSynchronize} calls. To address this, we implemented a custom fused and bucketized CUDA kernel for training-time AUC computation. By integrating the boolean mask directly into the kernel, we eliminated masking overhead and dynamic memory allocation, reducing metric update time from 66 ms to 0.5 ms per step while preserving metric accuracy (negligible AUC error).

\subsubsection{Optimizer Fusing and Gradient Scaling}
Optimizer inefficiencies were resolved by switching to a fused Adam implementation with integrated gradient scaling. Enabling the fused optimizer flag and incorporating the gradient scaler directly into the CUDA kernel removed redundant \texttt{cudaStreamSynchronize} calls, reducing optimizer step time from 40 ms to 20 ms.

\subsubsection{Fused Data Loading and Processing}
Previously, the data loading process involved significant latency ($\sim300$ms) per step due to input/output bottlenecks. To address the data loading bottleneck, we fused padding, batching, and packing operations into a C++ data loader to minimize Python multiprocessing overhead. These changes collectively reduced training step time by more than $50\%$.

\subsubsection{Parallelized Evaluation}
Previously, evaluation was performed in a serialized manner (once after several training steps), while evaluation is taking much less GPU memory compared with training, which causes GPU resource waste. In order to improve GPU efficiency, we save all of the checkpoints during training and run evaluation in parallel for all of the checkpoints at once. In this way, we load the dataset only once and run multiple forward passes in parallel to fully utilize the GPU memory, and reduced the overall runtime by 16\%.

\subsection{Analysis of energy consumption}
We analyze the relative energy consumption of {\systemname} compared to the previous production model. Results are shown in Table \ref{tab:energy}. The previous production model was trained on the same GPUs, but served online on CPUs. We can see that on a per-item basis {\systemname} is significantly more efficient than the previous system. We attribute this to effective compute amortization in both training and serving.
\begin{table}[htbp]
\begin{tabular}{@{}lll@{}}
\toprule
Energy Consumption Per Item & {\systemname} vs. Existing \\ \midrule
Training           & 0.2x             \\
Inference          & 0.7x             \\ \bottomrule
\end{tabular}
\caption{Relative energy consumption of {\systemname} compared to the previous production model during training and inference.}
\label{tab:energy}
\vspace{-2.5em}
\end{table}

\section{A/B Test Results}\label{sec:online-results}
Online results for {\systemname} are shown in Table~\ref{tab:online-results}. {\systemname} was evaluated in controlled online A/B tests against the previous production ranker, and now serves the majority of LinkedIn Feed traffic. Overall, {\systemname} improves time spent by +2.10\% and the volume of likes, comments, and reshares by +3.52\%. Broken down by member segment, gains are largest for the most active members, remain positive for less active members, and are neutral for new members. These online results exclude incremental training (Section~\ref{sec:incremental-training}).

\begin{table}[h]
\begin{tabular}{@{}llllll@{}}
\toprule
           & \multicolumn{5}{c}{Member Segment}                   \\ \midrule
           & Overall & DAU     & WAU     & MAU     & New          \\
Time spent & +2.10\% & +2.38\% & +1.84\% & +0.82\% & NSS \\
Like/Comment/Share & +3.52\% & +4.07\% & +3.40\% & +1.86\% & NSS \\\bottomrule
\end{tabular}

\caption{Online A/B test results of {\systemname} relative to the production model. All deltas are statistically significant unless marked NSS (not statistically significant).}
\label{tab:online-results}
\vspace{-2.5em}
\end{table}

\section{Deployment Lessons}
{\systemname} was a complete rewrite of the LinkedIn Feed ranking stack, spanning both modeling and infrastructure. Several key insights shaped the course of the project, which we discuss below.



\subsection{Offline–online alignment}\label{sec:offline-online-discrepancy}

A key tool for assessing online correctness was comparing offline and online scores on the same Feed SR sessions. We built a pipeline to score logged sessions offline and compare them with online scores, uncovering numerous online-stack bugs whose resolution narrowed the offline–online gap. Evaluation protocol also matters: unlike training, where negatives are downsampled, we evaluate on the full evaluation set. Downsampling negatives biases the evaluation distribution, inflates offline AUC, and weakens alignment with online metrics.

\subsection{Moving away from Java feature transformations}
Many of the issues from Section \ref{sec:offline-online-discrepancy} were due to offline/online feature discrepancies. Feature transformation was previously handled by shared Java-based transformations. With PyTorch, this shared framework no longer applies, requiring a new one to be built. While current online feature transformation is handled by a mix of Java and NumPy/PyTorch transformation, we are building a more principled Python-based shared feature transformation framework.

\section{Conclusion}
In this paper we have presented {\systemname}, a large-scale sequential recommendation model for LinkedIn's Feed. We have illustrated the modeling choices and system design that allowed us to create a better member experience than the previously existing production ranker. {\systemname} is now the majority member experience on LinkedIn's Feed.

\begin{acks}
This work represents the joint efforts across multiple teams. We would like to thank Deepak Agarwal, Tim Jurka, Balaji Krishnapuram, Xiaobing Xue, Hristo Danchev, Christine Lin, Xin Cai, Sudarshan Ramanujam, Shihai He for supporting this work.
\end{acks}

\bibliographystyle{ACM-Reference-Format}
\balance
\bibliography{base}

@inproceedings{wang2021dcn,
  title={Dcn v2: Improved deep \& cross network and practical lessons for web-scale learning to rank systems},
  author={Wang, Ruoxi and Shivanna, Rakesh and Cheng, Derek and Jain, Sagar and Lin, Dong and Hong, Lichan and Chi, Ed},
  booktitle={Proceedings of the web conference 2021},
  pages={1785--1797},
  year={2021}
}

@inproceedings{ma2018modeling,
  title={Modeling task relationships in multi-task learning with multi-gate mixture-of-experts},
  author={Ma, Jiaqi and Zhao, Zhe and Yi, Xinyang and Chen, Jilin and Hong, Lichan and Chi, Ed H},
  booktitle={Proceedings of the 24th ACM SIGKDD international conference on knowledge discovery \& data mining},
  pages={1930--1939},
  year={2018}
}

@inproceedings{xia2023transact,
  title={Transact: Transformer-based realtime user action model for recommendation at pinterest},
  author={Xia, Xue and Eksombatchai, Pong and Pancha, Nikil and Badani, Dhruvil Deven and Wang, Po-Wei and Gu, Neng and Joshi, Saurabh Vishwas and Farahpour, Nazanin and Zhang, Zhiyuan and Zhai, Andrew},
  booktitle={Proceedings of the 29th ACM SIGKDD Conference on Knowledge Discovery and Data Mining},
  pages={5249--5259},
  year={2023}
}

@article{zhai2024actions,
  title={Actions speak louder than words: Trillion-parameter sequential transducers for generative recommendations},
  author={Zhai, Jiaqi and Liao, Lucy and Liu, Xing and Wang, Yueming and Li, Rui and Cao, Xuan and Gao, Leon and Gong, Zhaojie and Gu, Fangda and He, Michael and others},
  journal={arXiv preprint arXiv:2402.17152},
  year={2024}
}

@inproceedings{borisyuk2024lirank,
  title={LiRank: Industrial Large Scale Ranking Models at LinkedIn},
  author={Borisyuk, Fedor and Zhou, Mingzhou and Song, Qingquan and Zhu, Siyu and Tiwana, Birjodh and Parameswaran, Ganesh and Dangi, Siddharth and Hertel, Lars and Xiao, Qiang Charles and Hou, Xiaochen and others},
  booktitle={Proceedings of the 30th ACM SIGKDD Conference on Knowledge Discovery and Data Mining},
  pages={4804--4815},
  year={2024}
}

@article{vaswani2017attention,
  title={Attention is all you need},
  author={Vaswani, Ashish and Shazeer, Noam and Parmar, Niki and Uszkoreit, Jakob and Jones, Llion and Gomez, Aidan N and Kaiser, {\L}ukasz and Polosukhin, Illia},
  journal={Advances in neural information processing systems},
  volume={30},
  year={2017}
}

@misc{Ghike_Gupta_2016, title={FollowFeed: Linkedin’s feed made faster and smarter}, url={https://www.linkedin.com/blog/engineering/feed/followfeed-linkedin-s-feed-made-faster-and-smarter}, journal={LinkedIn}, author={Ghike, Swapnil and Gupta, Shubham}, year={2016}, month={Mar}}

@article{ramanujam2025large,
  title={Large Scale Retrieval for the LinkedIn Feed using Causal Language Models},
  author={Ramanujam, Sudarshan Srinivasa and Alonso, Antonio and Kataria, Saurabh and Dangi, Siddharth and Gupta, Akhilesh and Tiwana, Birjodh Singh and Somaiya, Manas and Simon, Luke and Byrne, David and Ha, Sojeong and others},
  journal={arXiv preprint arXiv:2510.14223},
  year={2025}
}

@inproceedings{chen2019behavior,
  title={Behavior sequence transformer for e-commerce recommendation in alibaba},
  author={Chen, Qiwei and Zhao, Huan and Li, Wei and Huang, Pipei and Ou, Wenwu},
  booktitle={Proceedings of the 1st international workshop on deep learning practice for high-dimensional sparse data},
  pages={1--4},
  year={2019}
}

@inproceedings{zhou2018deep,
  title={Deep interest network for click-through rate prediction},
  author={Zhou, Guorui and Zhu, Xiaoqiang and Song, Chenru and Fan, Ying and Zhu, Han and Ma, Xiao and Yan, Yanghui and Jin, Junqi and Li, Han and Gai, Kun},
  booktitle={Proceedings of the 24th ACM SIGKDD international conference on knowledge discovery \& data mining},
  pages={1059--1068},
  year={2018}
}

@article{hertel2024efficient,
  title={Efficient user history modeling with amortized inference for deep learning recommendation models},
  author={Hertel, Lars and Daftary, Neil and Borisyuk, Fedor and Gupta, Aman and Mazumder, Rahul},
  journal={arXiv preprint arXiv:2412.06924},
  year={2024}
}

@article{naumov2019dlrm,
  title   = {Deep Learning Recommendation Model for Personalization and Recommendation Systems},
  author  = {Naumov, Maxim and Mudigere, Deepak and Shi, Hao-Jun Michael and Huang, Jianyu and Sundaraman, Narayanan and Park, Jongsoo and Wang, Xiaodong and Gupta, Udit and Wu, Carole-Jean and Azzolini, Alisson G. and others},
  journal = {arXiv preprint arXiv:1906.00091},
  year    = {2019},
  url     = {https://arxiv.org/abs/1906.00091}
}

@inproceedings{cheng2016wide,
  title     = {Wide \& Deep Learning for Recommender Systems},
  author    = {Cheng, Heng-Tze and Koc, Levent and Harmsen, Jeremiah and Shaked, Tal and Chandra, Tushar and Aradhye, Hrishi and Anderson, Glen and Corrado, Greg and Chai, Wei and Ispir, Mustafa and others},
  booktitle = {Proceedings of the 1st Workshop on Deep Learning for Recommender Systems (DLRS)},
  year      = {2016}
}

@inproceedings{wang2021dcnv2,
  title     = {{DCN} {V2}: Improved Deep \& Cross Network and Practical Lessons for Web-scale Learning to Rank Systems},
  author    = {Wang, Ruoxi and Fu, Bin and Fu, Gang and Wang, Ming},
  booktitle = {Proceedings of the Web Conference (WWW)},
  year      = {2021}
}

@inproceedings{ma2018mmoe,
  title     = {Modeling Task Relationships in Multi-task Learning with Multi-gate Mixture-of-Experts},
  author    = {Ma, Jiaqi and Zhao, Zhe and Yi, Xinyang and Chen, Jilin and Hong, Lichan and Chi, Ed H.},
  booktitle = {Proceedings of the 24th ACM SIGKDD International Conference on Knowledge Discovery \& Data Mining (KDD)},
  year      = {2018}
}

@inproceedings{kang2018sasrec,
  title     = {Self-Attentive Sequential Recommendation},
  author    = {Kang, Wang-Cheng and McAuley, Julian},
  booktitle = {Proceedings of the IEEE International Conference on Data Mining (ICDM)},
  year      = {2018},
  url       = {https://arxiv.org/abs/1808.09781}
}

@inproceedings{sun2019bert4rec,
  title     = {{BERT}4Rec: Sequential Recommendation with Bidirectional Encoder Representations from Transformer},
  author    = {Sun, Fei and Liu, Jun and Wu, Jian and Pei, Changhua and Lin, Xiao and Ou, Wenwu and Jiang, Peng},
  booktitle = {Proceedings of the 28th ACM International Conference on Information and Knowledge Management (CIKM)},
  year      = {2019},
  url       = {https://arxiv.org/abs/1904.06690}
}

@article{yang2025qwen3,
  title={Qwen3 technical report},
  author={Yang, An and Li, Anfeng and Yang, Baosong and Zhang, Beichen and Hui, Binyuan and Zheng, Bo and Yu, Bowen and Gao, Chang and Huang, Chengen and Lv, Chenxu and others},
  journal={arXiv preprint arXiv:2505.09388},
  year={2025}
}

@inproceedings{han2025mtgr,
  title     = {{MTGR}: Industrial-Scale Generative Recommendation Framework in Meituan},
  author    = {Han, Ruidong and Yin, Bin and Chen, Shangyu and Jiang, He and Jiang, Fei and Li, Xiang and Ma, Chi and Huang, Mincong and Li, Xiaoguang and Jing, Chunzhen and Han, Yueming and Zhou, Menglei and Yu, Lei and Liu, Chuan and Lin, Wei},
  booktitle = {Proceedings of the 34th ACM International Conference on Information and Knowledge Management (CIKM)},
  year      = {2025},
  url       = {https://arxiv.org/abs/2505.18654}
}

@article{deng2025onerec,
  title   = {{OneRec}: Unifying Retrieve and Rank with Generative Recommender and Iterative Preference Alignment},
  author  = {Deng, Jaixin and Wang, Shiyao and Li, Kuo Cai and Ren, Qigen and Hu, Qigen and Ding, Weifeng and Luo, Qiang and Zhou, Guorui},
  journal = {arXiv preprint arXiv:2502.18596},
  year    = {2025},
  url     = {https://arxiv.org/abs/2502.18596}
}

@article{loshchilov2017decoupled,
  title={Decoupled weight decay regularization},
  author={Loshchilov, Ilya and Hutter, Frank},
  journal={arXiv preprint arXiv:1711.05101},
  year={2017}
}

@article{su2024roformer,
  title={Roformer: Enhanced transformer with rotary position embedding},
  author={Su, Jianlin and Ahmed, Murtadha and Lu, Yu and Pan, Shengfeng and Bo, Wen and Liu, Yunfeng},
  journal={Neurocomputing},
  volume={568},
  pages={127063},
  year={2024},
  publisher={Elsevier}
}

@inproceedings{xiong2020layer,
  title={On layer normalization in the transformer architecture},
  author={Xiong, Ruibin and Yang, Yunchang and He, Di and Zheng, Kai and Zheng, Shuxin and Xing, Chen and Zhang, Huishuai and Lan, Yanyan and Wang, Liwei and Liu, Tieyan},
  booktitle={International conference on machine learning},
  pages={10524--10533},
  year={2020},
  organization={PMLR}
}

@inproceedings{touvron2021going,
  title={Going deeper with image transformers},
  author={Touvron, Hugo and Cord, Matthieu and Sablayrolles, Alexandre and Synnaeve, Gabriel and J{\'e}gou, Herv{\'e}},
  booktitle={Proceedings of the IEEE/CVF international conference on computer vision},
  pages={32--42},
  year={2021}
}

@inproceedings{bachlechner2021rezero,
  title={Rezero is all you need: Fast convergence at large depth},
  author={Bachlechner, Thomas and Majumder, Bodhisattwa Prasad and Mao, Henry and Cottrell, Gary and McAuley, Julian},
  booktitle={Uncertainty in Artificial Intelligence},
  pages={1352--1361},
  year={2021},
  organization={PMLR}
}

@article{chai2020highway,
  title={Highway transformer: Self-gating enhanced self-attentive networks},
  author={Chai, Yekun and Jin, Shuo and Hou, Xinwen},
  journal={arXiv preprint arXiv:2004.08178},
  year={2020}
}

@misc{casson2023transformerflops,
  author       = {Adam Casson},
  title        = {Transformer FLOPs},
  year         = {2023},
  howpublished = {\url{https://adamcasson.com/posts/transformer-flops}},
  note         = {Accessed: 2026-01-28}
}

@inproceedings{pancha2022pinnerformer,
  title={Pinnerformer: Sequence modeling for user representation at pinterest},
  author={Pancha, Nikil and Zhai, Andrew and Leskovec, Jure and Rosenberg, Charles},
  booktitle={Proceedings of the 28th ACM SIGKDD conference on knowledge discovery and data mining},
  pages={3702--3712},
  year={2022}
}


\end{document}